# Usages et conception des TIC : Proposition d'un modèle d'aide à la représentation de problème de conception

**Pierre Humbert,** Doctorant

**Equipe SITE** (LORIA / Université Nancy 2)
**Société Com-Médic**

LORIA / Equipe SITE
Campus Scientifique - BP 239
F-54506 Vandœuvre les Nancy

pierre.humbert@loria.fr

# 1. Introduction

Pour les entreprises, le succès d'un produit sur le marché est aujourd'hui devenu un véritable défi. Le nombre toujours croissant de demandes d'aide à l'innovation auprès d'organismes tels que OSEO montre que les idées ne sont pas absentes, les entreprises ne manquant pas de vanter les mérites et le caractère innovant des projets et des produits, qu'elles conçoivent. Cependant, le poids concurrentiel des grandes sociétés, la forte augmentation du nombre de création d'entreprises dans certains secteurs (OSEO, 2008) et un contexte technologique en perpétuelle évolution contribuent parfois à pousser les dirigeants de petites et moyennes entreprises dans des projets de conception qu'ils veulent rapides et fiables mais difficiles à maîtriser.

Cette conjoncture prend, sur la question de la conception de Technologies de l'Information et de la Communication (TIC), un relief particulier. Combien de dispositifs techniques, pourtant jugés innovants par leurs éditeurs concepteurs et prescripteurs, ne parviennent pas à être acceptés par les utilisateurs ? Sachant l'influence que peut avoir sur le dispositif conçu la manière dont les concepteurs se figurent les usagers (Flichy, 2008), les concepteurs donnent-ils suffisamment d'importance à la question de l'ancrage des dispositifs dans l'activité humaine et dans le contexte social au sein duquel ils sont immergés ? Le peuvent-ils et comment les y assister ? Telles sont les questions que nous examinerons dans cet article ou nous soulignerons tout d'abord l'existence d'un double processus de construction de sens entre conception et usage. Dans une seconde partie, les résultats d'une enquête sur les pratiques des éditeurs nous éclaireront sur les contraintes et approches de l'écoute des destinataires des TIC. Puis en montrant l'apport de l'intelligence économique comme processus de résolution de problème informationnel, nous proposerons un modèle d'aide à la conception qui s'en inspire.

# 2. Un double processus de construction de sens : usage et conception

L'innovation se caractérise par un fort degré d'incertitude que génère une double contrainte liée à l'originalité du produit ou service conçu et au succès de sa mise sur le marché. « L'innovation représente une invention qui a été acceptée par le marché » écrivent Prax et al. (2005 : 46). Pour assister l'entrepreneur innovateur dans sa tâche, la littérature regorge de modèles et méthodes vouées à l'aider dans sa démarche de conception innovante. Malgré cela, nombre de projets rencontrent des difficultés au contact de leurs destinataires. Une des raisons à cela réside, selon nous, dans la méconnaissance des concepteurs de la rencontre de deux processus, étudiée notamment par les sociologues de l'innovation, qui élaborent le sens du dispositif.

## 2.1. Le sens apporté au dispositif par les destinataires

L'équilibre entre ce qui est partagé et ce qui est propre à chacun de ces membres caractérise toute organisation au travers ses valeurs, ses pratiques, son vécu et ses finalités. L'introduction de TIC au sein d'un tel milieu doit immanquablement faire face à cet ensemble composite de perceptions individuelles et collectives. De nombreux travaux en sociologie et en sciences de l'information et de la communication, se sont intéressés au processus désigné par des termes tels qu'*appropriation*, *acceptation* ou *adoption*, d'un dispositif technique par les individus ou les groupes qui le reçoivent, l'utilisent ou l'achètent. Ainsi la Théorie de la diffusion de l'innovation de Rogers (1995) avance certains facteurs de succès de ce processus : Avantage relatif aux autres dispositifs, Compatibilité avec les valeurs, les expériences, les pratiques, les normes, etc., Complexité à comprendre et utiliser le dispositif, Testabilité, Observabilité des résultats et des gains attendus, Gain d'image ou de statuts.

Le Modèle d'acceptation des technologies (Davis et al., 1989) propose, quant à lui, deux points clés permettant de déterminer l'inclination d'un individu pour un dispositif :

- L'utilité perçue d'un système : l'accent est mis sur le rapport des technologies de l'information (TI) avec l'activité des utilisateurs. L'approche gestionnaire et cognitive des TI consiste à proposer des modèles de rapprochement entre les activités des utilisateurs au travail, les services et les contenus informationnels des systèmes et bases de données : modèle et profil utilisateur, métamodèle, etc.
- Sa facilité d'utilisation : l'accent est mis sur la relation entre l'utilisateur et le dispositif en situation d'usage, au travers des considérations d'ordre ergonomique et psychologique : analyse de tâches, exploitation de métaphores, etc.

D'autres études ont également porté sur la *non acceptation* au travers la notion de résistance aux changements. Ainsi, Brenot et Tuvée (1991, cités par Vas et Van de Velde, 2000) identifient plusieurs sources de résistances : Niveau de connaissance, d'éducation et d'ouverture d'esprit, Aptitudes au changement, Raisons économiques, Attitudes et préjugés, Craintes et conflits, sentiments de méfiance, pouvoir interpersonnel et intérêt.

Ces études avancent que le regard porté sur le dispositif est susceptible de lui faire incarner tantôt une menace, tantôt une opportunité (Desjardins et al. 2006). Ainsi, par exemple, l'adoption d'un dispositif aura comme effet de réveiller des conflits latents entre personnes ou groupes au sein d'une organisation (Vaillies, 2006 : 168) : enjeux d'identité et de pouvoir.

Nous pensons qu'au travers le processus d'appropriation, quel que soit le terme employé, le problème sous jacent consiste à examiner comment le destinataire fait sens de l'objet qu'on lui présente. Ainsi, comme l'écrit Akrich (1990 : 84), « *saisir la signification d'un dispositif technique c'est comprendre comment ce dispositif réorganise différemment le tissu de relations, de tout nature, dans lequel nous sommes pris et qui nous définissent* ». Une construction de sens qui rejoint le *sensemaking* défini par Weick (1995 : 4) comme une faculté d'assemblage à partir d'hypothèses, d'éléments extraits d'un vécu, d'autres représentations, au sein d'une représentation mentale unique qui crée de l'ordre et du sens.

Par conséquent, la question de l'appropriation est, selon nous, un processus de construction de sens que nombre de concepteurs ont parfois tendance à négliger. Ainsi, les phénomènes de résistance sont souvent considérés comme circonstanciels et détachés du dispositif, propres à l'organisation qui reçoit le dispositif.

## *2.2.    Le sens apporté au dispositif par les concepteurs*

Si la réception d'un dispositif technique implique que les destinataires fassent sens du dispositif, il n'en est pas moins vrai de la conception. Celle-ci désigne le processus qui part d'une analyse ou d'une connaissance préalable des besoins des destinataires pour arriver à la formulation des fonctions et caractéristiques du dispositif.

Dans ce processus, Flichy (2008 : 150) indique que « *les usages n'apparaissent qu'à travers les représentations que les concepteurs en ont* ». Akrich (1990) pour sa part, évoque l'élaboration de scripts ou scénarii qui définissent le cadre d'action des usagers, idée que nous retrouvons dans Woolgar (1991 : 71) pour lequel il est question d'une fiction sur les pratiques qui *configure* les usagers.

On peut également s'interroger sur l'influence des outils utilisés par les concepteurs, méthodes descriptives ou prescriptives (Dumas et al., 1990) ou objets intermédiaires. A travers leur formalisme (notamment graphique) et leur logique (héritage, hiérarchie, etc.) aide à constituer ce que l'on appelle un modèle, une représentation simplifiée d'une réalité et en cela sont des outils qui créent du sens.

Enfin, le processus de conception est un processus relevant d'une action collective. Il implique en effet un ensemble d'acteurs tels que chef de projet, dirigeant, développeur, expert du domaine, représentant d'utilisateurs, etc., aux connaissances et vécus divers, qui participent à l'élaboration du dispositif. Cette participation s'effectue immanquablement au travers la confrontation, la négociation de différentes perceptions, points de vue, représentations nécessaires à l'établissement d'un espace de compréhension mutuelle (Zarifian, 1998 : 16). En cela, le dispositif étant le produit de ces négociations, est doté du sens construit qui en résulte.

## *2.3. Intelligence économique et stratégie de conception*

La spécification des TIC peut être vue un processus d'acquisition de connaissances et de production d'information. Elle est également un processus décisionnel qui requière des acteurs-décideurs (tels que dirigeant, chef de projet, prescripteur, etc.) de faire des choix, de définir des orientations sur le dispositif, le discours d'accompagnement et l'approche du marché : L'implantation est-elle en bonne voie ? Quelles fonctions développer ? Quelles technologies exploiter ? Comment permettre une implantation réussie et durable ? Comment améliorer la perception du dispositif par les clients ?

Nous faisons l'hypothèse qu'une démarche d'intelligence économique (IE) peut contribuer au processus de conception. L'IE peut être définie comme un « *processus de collecte, de traitement et de diffusion de l'information qui a pour objet la réduction de la part d'incertitude dans la prise de toute décision stratégique* » (Revelli, 1998).

David (2005) définit le processus d'IE au travers 7 étapes : (1) Définition du problème décisionnel, (2) Transformation du problème décisionnel en problème de recherche d'information, (3) Identification de sources, (4) Recherche d'information, (5) Traitement, analyse et fourniture de l'information, (6) Interprétation des informations collectées et (7) Décision. Selon ce processus peut contribuer à la conception des TIC en :

- Proposant une méthodologie simple de résolution de problèmes orientée décision,
- Aidant à expliciter et comprendre un besoin, un point de vue sur l'objet du problème,
- Sensibilisant au recensement de multiples sources d'information,
- Guidant le recueil d'informations,
- Proposant des outils d'analyse d'information.

Comme nous le verrons au travers les résultats de notre enquête, cette démarche nous semble particulièrement pertinente pour des petits éditeurs, acteurs essentiels du développement des TIC, confrontés à un environnement changeant et parfois semblant peu armés face à l'incertitude qui caractérise ce type de projet.

## 3. Les pratiques de conception dans les PME

Comme nous l'avons vu, la question de l'acceptation des dispositifs techniques par les destinataires est un problème essentiel pour les auteurs de ces dispositifs. Or, bien que les TIC constituent une grande part des projets d'innovation des petites entreprises (40% des projets de R & D (DGRI, 2008 : 44)), il semble que nous ne connaissons pas grand-chose de leurs méthodes. En effet, hormis quelques exceptions (par exemple au Canada : Aranda et al., 2007), la littérature portant sur l'ingénierie des besoins et les méthodes de conception en générale évoque majoritairement des cas de grandes entreprises, comportant souvent des services R & D et possédant des compétences et des ressources adéquates. Or, par rapport à ces dernières, les petites structures, diffèrent par leur organisation et sont souvent plus exposées aux aléas de leur environnement tout en ne possédant pas nécessairement les mêmes ressources ni les mêmes outils.

Devant ce constat, nous nous sommes interrogés sur leurs capacités à exploiter les méthodes et outils de conception préconisés dans les manuels. Ont-elles les moyens nécessaires pour conduire ces méthodes ? Leur sont-elles un secours ? Existe-t-il des phénomènes d'adaptation et quels sont les facteurs qui les poussent à s'adapter ? Et enfin, dans un tel contexte, les concepteurs ont-ils les capacités d'ouvrir le niveau de leur analyse à des paramètres autres que techniques ? Aussi avons-nous cherché à vérifier si les petites entreprises emploient des méthodes bien définies, ou si elles s'adaptent en fonction notamment de leur morphologie et de leur contexte, pratiquant une forme de « bricolage organisé » (Vacher, 2006).

### 3.1. Méthodologie

Pour vérifier cette hypothèse, nous avons tenté d'analyser plusieurs cas. L'un d'eux résulte de notre propre expérience de l'entreprise en convention CIFRE, cadre de notre thèse, au sein d'une petite société d'édition médicale, instigatrice de la conception d'un outil collaboratif à destination des professionnels de l'accompagnement de la personne handicapée. Ce premier cadre a constitué la base de nos questions et de ce que nous avons cherché à vérifier par une enquête réalisée auprès de petites entreprises dont l'activité principale est de concevoir et distribuer des logiciels.

A partir des fichiers de Chambre de Commerce et d'Industrie (Lorraine et Alsace, principalement), nous avons constitué un échantillon composé de 25 petites sociétés, répondant aux critères de faible effectif et d'activité de conception et commercialisation de logiciels. Nous avons mené l'enquête par questionnaires envoyés par courriel et quand cela était possible, une rencontre avec le responsable de l'entreprise était également organisée. Les questions se présentaient selon plusieurs axes :

- Entreprise : effectif, produit, stade de conception, etc.
- Acteurs : Qui participe à la conception ? Cette participation se cumule-t-elle avec d'autres fonctions ?
- Méthodes : Quelles méthodes utilisez-vous pour recueillir les besoins des destinataires ? En êtes-vous satisfait ? Avez-vous développé une méthode particulière ?

- Intérêt pour le social : Vous sentez-vous concernés par les phénomènes de résistance au changement ?
- Pratiques informationnelles : Comment recueillez-vous ces informations ? Les conservez-vous ? De quelle manière ? Comment les exploitez-vous ?

Les retours ont été assez peu nombreux : malgré nos relances, seul 20% des organisations contactées ont répondu, parfois de manière incomplète.

Conscient que ceci limite la portée de notre propos, nous pensons cependant que les résultats recueillis nous permettent de dégager des tendances intéressantes qui tendent à vérifier notre intuition sur le sujet.

## *3.2. Des facteurs internes et externes d'adaptation*

Dès les premiers résultats de notre étude il semble qu'élaborer un dispositif recouvre des réalités très différentes selon les entreprises. Les méthodes et modèles existants sont pour elles des cadres conventionnels, considérés parfois comme nécessaires (contractualisation avec le client, image de l'entreprise, etc.) mais particulièrement lourds à mettre en œuvre (documentation fréquente, procédures rigides, etc.). Tels quels, ces cadres méthodologiques et normatifs, bien que servant de référence pour qui les connait, sont souvent jugés inadaptés au contexte et à la morphologie d'une petite structure.

### 3.2.1. La familiarité du dirigeant avec ces méthodes influence leur adoption

Il apparaît au terme de notre enquête que l'exploitation de méthodes n'est pas systématique : il semble que seules les entreprises dont le fondateur possède une expérience de la méthodologie (au cours d'une vie professionnelle antérieure par exemple), mettront en œuvre une démarche qui s'en inspire. De plus, le profil du dirigeant semble influencer particulièrement l'adoption d'une telle démarche. En effet, alors que les membres d'une entreprise (tels que chef de projet ou développeur) connaissent des méthodes, si le dirigeant ou fondateur ne possède pas cette connaissance, son apport méthodologique pourra être mal pris en considération.

### 3.2.2. Inadaptation des méthodes avec le fonctionnement d'une petite entreprise

La seconde caractéristique est triple et d'ordre organisationnel. Elle englobe les aspects liés à l'effectif de l'entreprise, au temps et à la polyvalence des rôles des acteurs. En effet, les effectifs réduits des entreprises interrogées (de 3 à 8 salariés) permettent rarement une spécialisation sur des projets ou fonctions. Il est difficile pour chacun, compte tenu des ressources disponibles impliquant une certaine polyvalence dans les activités, de respecter la temporalité qu'impose une méthodologie.

### 3.2.3. Incompatibilités des méthodes avec la représentation du client du service

Si l'utilisation de méthodes constitue un coût humain et temporel pour l'entreprise (on peut imaginer qu'il puisse être également financier, à terme), il constitue aussi un coût pour le client, dans le cas d'une offre orientée client. En effet, le temps qui lui est facturé, consacré à la modélisation et la formalisation préalable d'un besoin (comme le préconise notamment la démarche UML) peut être jugée par le client comme du temps improductif et donc non nécessaire.

### 3.2.4. Un cadre d'échange sur le mode informel

L'une des fonctions des méthodes de conception consiste à rationaliser et fournir un cadre aux échanges, d'une part entre les concepteurs et les destinataires, d'autre part entre concepteurs (fonction de traçabilité, mémorisation, etc.). Cette fonction présente un grand intérêt pour les grandes sociétés souvent segmentée sur le plan organisationnel (en services) ou sur le plan géographique (délocalisation ou filiales). Dans le cas des petites entreprises, la situation est tout autre car la segmentation est beaucoup moins marquée et la proximité des acteurs renforcée. Ainsi, les échanges se réalisent davantage sur le mode informel, il est plus facile de discuter les options et de négocier les points de vue dans un tel cadre.

## *3.3. L'écoute des besoins de l'utilisateur et du client*

Notre enquête a également pointé quelques particularités quant aux rapports des concepteurs avec leurs clients et l'écoute qu'ils doivent mettre en œuvre. Nous avons interrogé les acteurs afin de savoir qui recueillait ces besoins, ce qu'ils recueillaient et comment ils l'exploitaient.

### 3.3.1. Une représentation empirique de la cible

Le rapport entre dirigeant et cible de l'entreprise influence l'écoute apportée. En effet, nombre de dispositifs développés par des petites sociétés, se fondent sur un besoin identifié sur le terrain par des acteurs ayant auparavant eu une proximité avec celui-ci (métier ou fonction antérieure). Ainsi, ceux-ci s'appuient sur leurs connaissances pratiques des problématiques, du domaine, des activités, etc. Dans ce cas, relativement à la confiance que possèdent les acteurs en cette expérience, l'utilisation de méthodologies n'est pas d'un apport essentiel selon eux, ils préféreront accorder davantage d'importance à leur intuition et à leur représentation du domaine ciblé. Ce point est, à notre sens, un piège réel dans la mesure où il filtre la réalité des destinataires en fonction du vécu et des connaissances du concepteur.

### 3.3.2. Le double rôle du dirigeant

Dans de nombreux cas, le fondateur - dirigeant joue le rôle de l'analyste, parfois accompagné d'un technicien. La place du dirigeant ici se justifie par le besoin d'assurer la relation avec le client qui dépasse les considérations relatives à la conception du dispositif (par exemple : besoin de créer un climat de confiance). Cette double fonction comporte un risque également, celui de privilégier la relation au détriment d'une démarche approfondie d'écoute des besoins.

### 3.3.3. Le choix circonstanciel de l'analyste

Un des cas rencontré montre que l'analyste n'est pas toujours la même personne, ni la plus apte à le faire. En effet, pour ne pas ralentir l'activité de l'entreprise, les critères qui déterminent la personne qui recueillera les besoins du client sont souvent liés à la disponibilité de chaque acteur, de la charge de travail en cours, etc. plutôt que par sa compétence à le faire ou sa connaissance du domaine.

### 3.3.4. Le contexte social peu considéré

Pour tenter de rechercher comment les petites entreprises prenaient en compte le tissu social et l'impact que pouvait avoir le dispositif technique sur ce tissu, nous nous sommes intéressés à leur méthode d'appréhension des phénomènes de résistance au changement. Notons que tous les dispositifs dont il était question dans notre enquête n'apportent pas les changements significatifs à l'origine de ces résistances. Dans les cas susceptibles d'en générer, les concepteurs interrogés, bien que conscients des risques qu'ils courent vis-à-vis du succès de leur dispositif dans ce contexte, ne disposent pas d'outils ou méthodes dédiées à ces phénomènes. Il est couramment admis que ce type de problème relève plus de la compétence du client que du concepteur du dispositif, réduisant ainsi ces incidents à des considérations managériales sans remise en cause des choix de conception.

Selon notre enquête, proposer une méthode aux petits éditeurs requière la prise en compte de certains critères d'acceptation tel que le coût temporel, la charge de travail que cela représente, la sensibilisation du dirigeant à une démarche structurée et la disposition de ce dernier à remettre en question ses *a priori* sur les destinataires. Aussi avons-nous tenté de prendre en compte ces conditions dans notre proposition qui se veut être une démarche simple (s'opposant ainsi à la complexification croissante des méthodes proposées), structurée (fournissant un cadre, un guide) et ouverte (sur les infinies possibilités qui caractérise à la fois un environnement instable et la démarche d'innovation elle-même).

## 4. Proposition d'un modèle d'aide à la représentation d'un problème de conception

La question de l'écoute des destinataires est une démarche stratégique cruciale en situation d'innovation. Nous avons vu que l'adoption de TIC au sein d'organisations est un processus complexe qui requière, selon nous, un minium de méthode. Or, d'après notre enquête, une méthodologie ne sera acceptée par les éditeurs qu'à condition qu'elle ne remplisse certaines conditions. Cette situation a motivé notre proposition d'un processus aidant la représentation d'un problème de conception.

La démarche méthodologique que nous proposons est simple : confrontées à une situation problématique, les concepteurs et les destinataires des TIC, sont guidés afin de faire émerger leur perception d'un même objet : le dispositif qui, selon nous, les relie. Les perceptions ou représentations, sont ensuite comparées afin de tenter d'identifier les conflits, les incohérences, les malentendus, etc. Les résultats de cette comparaison sont ensuite exploités pour connaître les points sur lesquels il peut être pertinent d'agir et préparer les choix d'action à travers des propositions de plan d'action. Nous proposons une démarche en 4 temps : (1) Analyse de la situation, (2) Identification d'un problème ou formulation d'une hypothèse, (3) Comparaison des représentations pour vérifier cette hypothèse et (4) Proposition de plan d'action.

Selon le processus d'IE, la compréhension d'un problème débute par l'identification des enjeux, notion approfondie par Kislin (2007), et l'appréciation générale d'une situation au travers l'examen des signaux révélant la présence d'un problème. La première phase réalise en partie cette tâche en combinant deux modalités d'observation, l'une active, l'autre passive. Cette dernière consiste en une veille constante des facteurs de satisfaction des destinataires. Elle exploite des sources externes (remontées des utilisateurs, clients, etc.) et internes (testeurs, experts, etc.). La modalité active se rapporte plutôt à un travail d'exploration, d'investigation portant sur leurs représentations sociales et individuelles (problème à résoudre, solutions envisagées, ressenti, influence souhaitée/subie du dispositif, attentes techniques, organisationnelles et sociales, etc.). Des techniques d'analogies rationnelles (Aznar, 2005 : 107), conformes aux exigences d'ouverture et de structure vues plus haut, amèneraient les individus à croiser des éléments issus de leur imaginaire (pouvoir, savoir, espace, temps, social, gain, équité, etc.). De plus, l'opérateur de pensée développé par Goria (2006) fait émerger les perceptions d'un objet au travers trois perspectives : la structure ontologie de l'objet, sa temporalité et son but, il constitue pour nous un outil tout à fait pertinent.

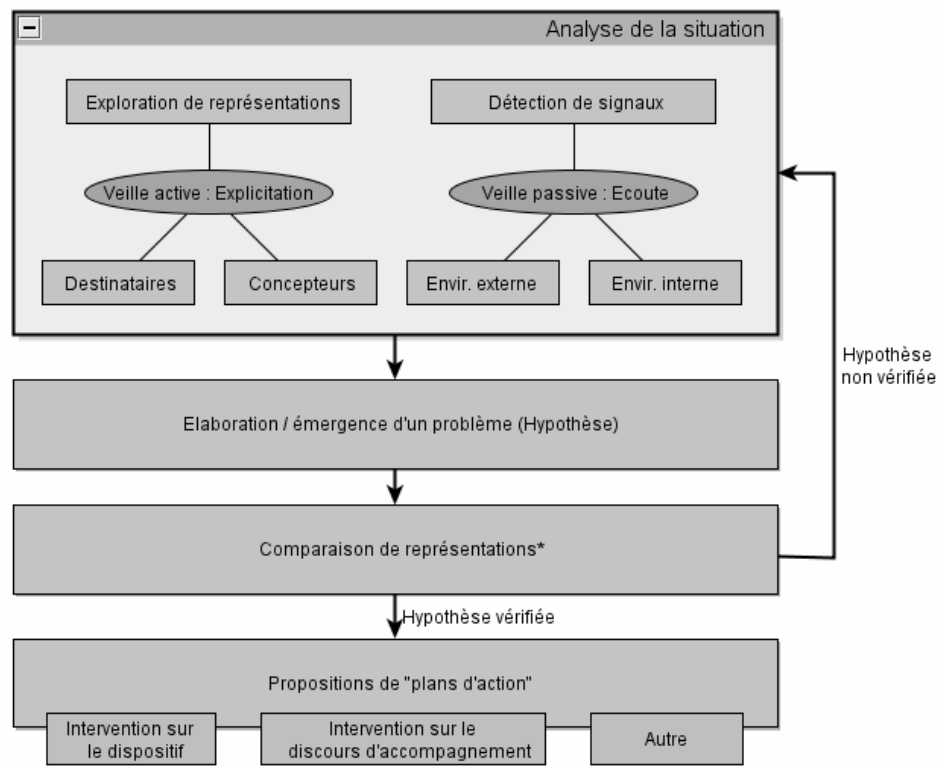

**Figure 1 : représentation du modèle proposé**

Nous pensons que, par défaut, la première étape est initiée par une intention préventive (*au cas où…*) ou exploratoire (*vérifions si…*), elle ne fait donc pas partie du processus de résolution de problème. La seconde étape permet de rentrer dans ce processus en construisant un problème à partir des signaux captés lors de la première phase d'analyse. Ce problème n'est alors perçu que sous la forme d'une hypothèse que l'étape suivante va tenter de valider ou d'invalider.

La troisième étape vise à comparer les représentations des destinataires et des concepteurs, recueillies lors de l'analyse de la situation. Sur ce point, nous ne proposons pas encore de méthode qui convienne à notre cadre particulier. Citons cependant les travaux d'Easterbrook (1991) qui aborde précisément cette question pour la spécification d'applications informatiques. Il propose la représentation logique de points de vue recueillis et des algorithmes permettant de confronter les propositions et détecter les points d'accord ou de désaccord entre plusieurs individus interrogés. Sa démarche, sans doute trop complexe vis-à-vis de notre cadre d'application, constitue une piste intéressante.

Enfin, la quatrième phase part des résultats de la phase précédente : les incohérences, les écarts et malentendus sur les attentes exprimées, supposées ou analysées au démarrage du projet. Il est possible alors, en considérant la situation appréciée au début du processus, de proposer des pistes d'action, sur le plan technique en re-conception ou reconsidération de l'offre proposée aux clients / utilisateurs, sur le plan social, champ d'action relevant soit du client (Dimension managériale) soit des concepteurs (Quels sont les points techniques générant une influence négative sur les rapports sociaux des acteurs ?) et enfin sur le plan de la communication relevant des réalisateurs, en particulier commerciaux et dirigeant, qui peuvent être amenés à remettre en question le discours d'accompagnement du dispositif technique.

## 5. Conclusion

Nous avons tenté de mettre en avant la dimension stratégique du changement duquel émergent des appréhensions, des attentes inexprimées jusqu'alors. De là nous avons tenté de montrer l'apport d'une démarche d'écoute fondée sur le processus d'intelligence économique dans l'orientation d'un projet de conception de TIC. Des méthodes existent dans ce domaine (qualité, design, gestion de projet, etc.) mais nous avons pu entrevoir l'écart qu'il existait, au sein des petites entreprises, entre les pratiques prescrites et la réalité. Aussi avons-nous formulé une proposition d'un modèle d'aide à la conception qui fournit aux acteurs un outil simple de recueil et d'analyse destiné à évaluer la stratégie de conception du dispositif au contact de ses destinataires.

## 6. Bibliographie